\documentclass[conference]{IEEEtran}
\IEEEoverridecommandlockouts
\usepackage[letterpaper, top=0.75in, bottom=0.96in, left=0.660in, right=0.680in]{geometry}
\usepackage{cite}
\usepackage{bm}
\usepackage{amsmath,amssymb,amsfonts}
\usepackage{algorithmic}
\usepackage{graphicx}
\usepackage{textcomp}
\usepackage{xcolor}
\usepackage{cases}
\usepackage{algorithm}
\usepackage{url}
\usepackage{float}
\usepackage{color}

\renewcommand{\algorithmicrequire}{ \textbf{Input:}}     
\renewcommand{\algorithmicensure}{ \textbf{Output:}}    

\newtheorem{proposition}{\underline{Proposition}}

\def\BibTeX{{\rm B\kern-.05em{\sc i\kern-.025em b}\kern-.08em
		T\kern-.1667em\lower.7ex\hbox{E}\kern-.125emX}}
\begin{document}
	
	\title{Optimal Transmit Signal Design for Multi-Target MIMO Sensing Exploiting Prior Information}
	\author{\IEEEauthorblockN{Jiayi Yao and Shuowen Zhang}
		\IEEEauthorblockA{Department of Electrical and Electronic Engineering, The Hong Kong Polytechnic University\\
			E-mail: jiayi.yao@connect.polyu.hk, shuowen.zhang@polyu.edu.hk}}
	\maketitle
\addtolength{\topmargin}{0.06in}
	\begin{abstract}
		In this paper, we study the transmit signal optimization in a multiple-input multiple-output (MIMO) radar system for sensing the angle information of \emph{multiple targets} via their reflected echo signals. We consider a challenging and practical scenario where the angles to be sensed are \emph{unknown} and \emph{random}, while their probability information is known \emph{a priori} for exploitation. First, we establish an analytical framework to quantify the multi-target sensing performance exploiting prior distribution information, by deriving the \emph{posterior Cram\'er-Rao bound (PCRB)} as a lower bound of the mean-squared error (MSE) matrix in sensing multiple unknown and random angles. Then, we formulate and study the transmit sample covariance matrix optimization problem to minimize the PCRB for the sum MSE in estimating all angles. Moreover, we propose a sum-of-ratios iterative algorithm which can obtain the optimal solution to the PCRB-minimization problem with low complexity. Numerical results validate our results and the superiority of our proposed design over benchmark schemes.
	\end{abstract}
	
	\section{Introduction}
\addtolength{\topmargin}{-.1in}
	Multiple-input multiple-output (MIMO) radar is widely known for its ability of offering superior waveform diversity and design flexibility with smart antenna techniques \cite{fishler2004mimo}. To unleash the full potential of MIMO radar, the transmit signals need to be judiciously designed. Along this line, the vast majority of existing works studied the transmit signal or beamforming optimization for localizing targets assuming the exact or approximate location parameters of the targets are known. Under this assumption, a lower bound of the mean-squared error (MSE) termed as $\textit{Cram\'er-Rao bound (CRB)}$ \cite{van2004detection,forsythe2005waveform,bekkerman2006target,li2007range,liu2021cramer,hua2023mimo} was typically adopted as the performance metric. For instance, CRB was first introduced in \cite{forsythe2005waveform} to MIMO radar for clutter-free angle estimation in narrowband systems. Considering several design criteria, such as the trace, determinant, and the largest eigenvalue of the CRB matrix, \cite{li2007range} studied several MIMO radar waveform optimization problems in various scenarios. Recently, integrated sensing and communication (ISAC) has attracted significant research interests, where CRBs for both point target and extended target were derived for guiding the transmit signal optimization while guaranteeing a required communication quality at the users \cite{liu2021cramer}. 
	
	However, in practice, the location parameters to be sensed can be \emph{unknown} and \emph{random}, while the distribution of them can be known \emph{a priori} based on e.g., target appearance pattern and statistical information. In this case, 
	$\textit{posterior Cram\'er-Rao bound (PCRB)}$ or Bayesian Cram\'er-Rao bound (BCRB) can be adopted to characterize the lower bound of the MSE when the $\textit{prior}$ \emph{distribution information} of the $\textit{unknown}$ and $\textit{random}$ parameters is exploited \cite{attiah2023active,xu2023mimo,xu2023mimo1,xu2024integrated,hou2023secure,hou2023optimal}. \cite{attiah2023active} studied the Bayesian sequential beamforming optimization in an ISAC system. \cite{xu2023mimo} studied a MIMO radar system which aims to estimate the angle information of a point target. It was shown that the PCRB-minimizing transmit signal design will result in a novel \emph{probability-dependent power focusing} effect. This work was extended to a MIMO ISAC system, where useful properties on the optimal transmit covariance matrix were derived \cite{xu2023mimo1}. \cite{hou2023secure,hou2023optimal} considered a secure ISAC system where a multi-antenna base station (BS) communicates with one user and senses the location of a potential eavesdropping target simultaneously. \cite{du2024uav} extended the exploitation of prior distribution information to the trajectory optimization of a sensing drone. Nevertheless, existing works along this line were focused on the sensing of a \emph{single} target. With multiple targets, the optimal transmit signal design for optimizing the overall sensing performance of all targets still remains an open problem; moreover, how to exploit the prior joint distribution of all targets' parameters in the design is also challenging. This thus motivates our study in this paper.
	
	\begin{figure}[t]
		\centering
		\setlength{\abovecaptionskip}{-0.2cm}
		\includegraphics[width=0.36\textwidth]{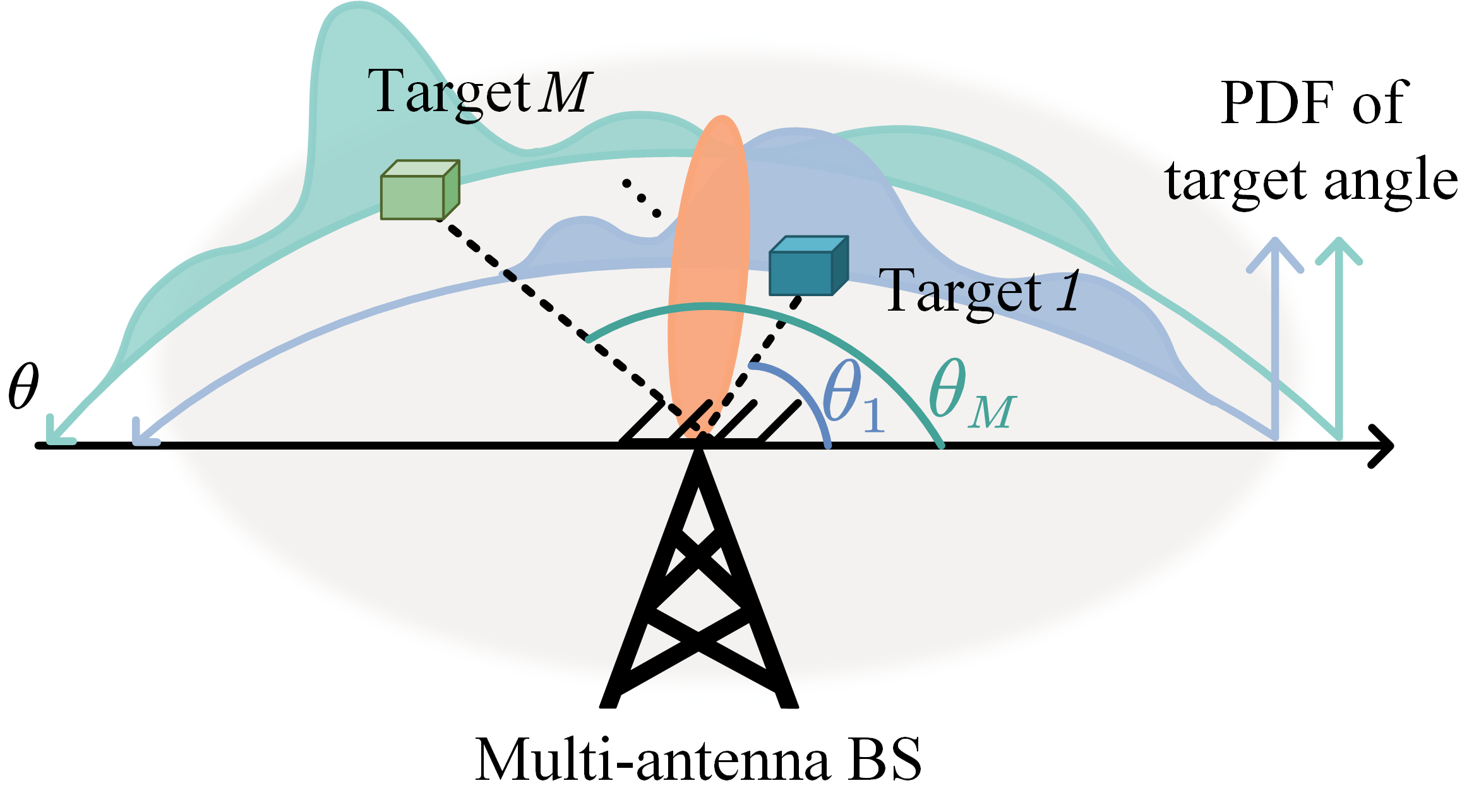}
		\caption{Illustration of multi-target MIMO sensing with prior information.}\label{Fig_System_Model}
		\vspace{-0.88cm}
	\end{figure}
		
	We consider a MIMO radar system with multiple targets, whose \emph{unknown} and \emph{random} angles need to be sensed by exploiting their prior distribution information. To quantify the multi-target sensing performance, we derive the PCRB for the sensing MSE matrix as an explicit function of the transmit sample covariance matrix, which is general for any prior distribution. Based on this, we formulate the transmit sample covariance matrix optimization problem to minimize the PCRB for the sum MSE in sensing all targets' angles, which is a convex optimization problem. Moreover, we propose a novel sum-of-ratios iterative algorithm to obtain the optimal solution with low complexity. It is shown via numerical results that our proposed design outperforms various benchmark designs.
	\section{System Model}
\addtolength{\topmargin}{-.1in}
	Consider a MIMO radar system with $N_t\!\geq\! 1$ transmit antennas and $N_r\!\!\geq\!\! 1$ co-located receive antennas at the BS, which aims to sense the \emph{unknown} and \emph{random} angle information of $M\!\!\geq\!\! 1$ point targets via the echo signals reflected by the targets. Let $\theta_m$ denote the angle of each $m$-th target with respect to the BS, as illustrated in Fig. \ref{Fig_System_Model}. Let $p_{\boldsymbol{\Theta}}(\boldsymbol{\theta})$ denote the joint probability density function (PDF) of all angles in $\boldsymbol{\theta}=[\theta_1,...,\theta_M]^T$, which is assumed to be known \emph{a priori} via target appearance pattern or statistical information \cite{xu2023mimo,xu2023mimo1,xu2024integrated,hou2023optimal,hou2023secure}.
	
	To unveil fundamental insights on multi-target sensing exploiting prior information, we focus on the case where the channel between the BS and each target is a line-of-sight (LoS) channel. The channel from the BS transmitter to the BS receiver via the reflection of the $m$-th target is modeled as
	\vspace{-0.35cm}
	\begin{equation}
		\boldsymbol{G}_m(\theta_m) =\alpha_m\boldsymbol{b}(\theta_m) \boldsymbol{a}^H(\theta_m),\quad m=1,...,M.\label{channel_m}
     \vspace{-0.15cm}
	\end{equation}
	Specifically, $\alpha_m=\frac{\beta_0}{r_{m}^{2}}\psi_m\in \mathbb{C}$ denotes the overall reflection coefficient for the $m$-th target, where $\beta_0$ denotes the reference channel power at distance $1$ meter (m), $r_m$ denotes the distance between the BS and the $m$-th target in m, and $\psi_m\in \mathbb{C}$ denotes the radar cross section (RCS) coefficient for the $m$-th target. The reflection coefficient is \emph{unknown} since the RCS coefficient $\psi_m$ is unknown. We further consider the case where $\alpha_m$ is a \emph{random} parameter following the circularly symmetric complex Gaussian (CSCG) distribution with zero mean and variance $2\sigma_{\alpha_m}^2$ \cite{attiah2023active}, which is known \emph{a priori};\footnote{Note that this assumption holds for various scenarios, e.g., when $r_m$ is known and $\psi_m$ is a complex Gaussian random variable. Moreover, our results can be readily extended to the case with other distributions for $\alpha_m$'s.} while $\alpha_m$'s for different targets are independent of each other and the targets' angles $\theta_m$'s. Moreover, $\boldsymbol{b}(\theta_m)\in \mathbb{C}^{N_r\times 1}$ denotes the array steering vector at the BS receive antennas for angle-of-arrival $\theta_m$, and $\boldsymbol{a}(\theta_m)\in \mathbb{C}^{N_t\times 1}$ denotes the array steering vector at the BS transmit antennas for angle-of-departure $\theta_m$. We assume all the $M$ targets are within the same range bin, thus the overall multi-target MIMO reflection channel is given by $\sum_{m=1}^M\boldsymbol{G}_m(\theta_m)$.
	
	Denote $L\geq 1$ as the total number of samples of the transmit probing signal for sensing $\boldsymbol{\theta}$. Let $\boldsymbol{x}_l\in \mathbb{C}^{N_t\times 1}$ denote the baseband equivalent transmit probing signal vector at the $l$-th sample, and $\boldsymbol{X}=[\boldsymbol{x}_1,...,\boldsymbol{x}_L]\in\mathbb{C}^{N_t\times L}$ denote the collection of all probing signal vectors. Let $\boldsymbol{R}_X=\frac{1}{L}\sum_{l=1}^L{\boldsymbol{x}_l\boldsymbol{x}_{l}^{H}}=\frac{1}{L}\boldsymbol{XX}^H$ denote the transmit sample covariance matrix. Denote $P$ as the total transmit power budget, which yields $\mathrm{tr}\left( \boldsymbol{R}_X \right) \leq P$. Let $\boldsymbol{n}_l\in \mathbb{C}^{N_r\times 1}$ denote the CSCG noise vector at the BS receiver for the $l$-th sample, where $\boldsymbol{n}_l\sim \mathcal{C}\mathcal{N}(\mathbf{0},\sigma^2\boldsymbol{I}_{N_r})$ with $\sigma^2$ denoting the average receiver noise power. Define $\boldsymbol{N}=[\boldsymbol{n}_1,...,\boldsymbol{n}_L]$. Let $\boldsymbol{y}_l\in \mathbb{C}^{N_r\times 1}$ denote the echo signal vector of the transmit probing signal which is reflected by the targets and received back at the BS receive antennas. The collection of all received echo signals $\boldsymbol{Y}=[\boldsymbol{y}_1,...,\boldsymbol{y}_L]\in \mathbb{C}^{N_r\times L}$ is thus given by
	\vspace{-0.1cm}
	\begin{align}
		\boldsymbol{Y}\!\!=\!\!\sum_{m=1}^M\boldsymbol{G}_m(\theta_m) \boldsymbol{X}\!\!+\!\!\boldsymbol{N}\!\!=\!\!\sum_{m=1}^M\alpha_m\boldsymbol{b}(\theta_m) \boldsymbol{a}^H(\theta_m)\boldsymbol{X}\!\!+\!\!\boldsymbol{N}.	\label{Y}
	\end{align}
	
	Note that the received echo signals in $\bm{Y}$ involve both the unknown location parameters to be sensed, $\bm{\theta}$, and another set of unknown parameters, $\{\alpha_m\}_{m=1}^M$. Thus, to obtain an accurate estimate of $\bm{\theta}$, $\bm{\theta}$ and $\{\alpha_m\}_{m=1}^M$ need to be jointly sensed based on $\bm{Y}$ and the prior distribution information of $\bm{\theta}$ and $\{\alpha_m\}_{m=1}^M$. In the following, we first characterize the multi-target sensing performance for $\bm{\theta}$ exploiting prior distribution information as an explicit function of the transmit signal design; then, we study the transmit signal design to optimize the sensing performance for $\bm{\theta}$.
	
	\section{Multi-Target Sensing Performance Characterization Exploiting Prior Information}
	In this section, we aim to characterize the multi-target sensing performance for $\bm{\theta}$ when $\bm{\theta}$ and $\{\alpha_m\}_{m=1}^M$ are jointly estimated. Specifically, since the sensing MSE is difficult to be analytically expressed especially for the multi-target case, we propose to derive the PCRB as a lower bound of the MSE when prior distribution information is available. Note that the PCRB is generally tight in the moderate-to-high signal-to-noise ratio (SNR) regime \cite{van2004detection}, thus being a suitable performance metric in high-accuracy sensing.
	
	Define $\boldsymbol{\zeta }=[\boldsymbol{\theta}^T,\boldsymbol{\alpha }^{T}] ^T\in \mathbb{R} ^{3M \times 1}$ as the collection of all unknown parameters, where $\boldsymbol{\alpha}=[\alpha _{1}^{\mathrm{R}},\alpha_1^{\mathrm{I}},...,\alpha_M^{\mathrm{R}},\alpha_M^{\mathrm{I}}] ^T$ with $\alpha_m^{\mathrm{R}}$ and  $\alpha_m^{\mathrm{I}}$ denoting the real and imaginary parts of $\alpha_m$, respectively. Note that since $\alpha_m\sim \mathcal{CN}(0,2\sigma_{\alpha_m}^2)$,  $\alpha_m^{\mathrm{R}}$ and $\alpha_m^{\mathrm{I}}$ are independent zero-mean Gaussian random variables with variance $\sigma_{\alpha_m}^2$, whose PDF is given by $p_{\alpha_m}(\bar{\alpha}_m)=e^{-\frac{\bar{\alpha}_m^2}{2\sigma_{\alpha_m}^{2}}}/\sqrt{2\pi}\sigma_{\alpha_m},\  \bar{\alpha}_m\in \{\alpha_m^{\mathrm{R}},\alpha_m^{\mathrm{I}}\}$. By further noting that all the parameters in $\bm{\zeta}$ are independent of each other, the PDF of $\boldsymbol{\zeta}$ is given by 
	\vspace{-0.3cm}
	\begin{align}\label{p_z}
		p_{\mathrm{Z}}(\boldsymbol{\zeta})=p_{\bm{\Theta}}(\boldsymbol{\theta})\prod_{m=1}^M{p_{\alpha_m}(\alpha_m^{\mathrm{R}})p_{\alpha_m}(\alpha_m^{\mathrm{I}})}.
	\end{align}
	\vspace{-0.25cm}
	
	With the above prior information available for exploitation, the Fisher information matrix (FIM) is given by $\bm{F}=\bm{F}_{\mathrm{O}}+\bm{F}_{\mathrm{P}}$ \cite{shen2010fundamental}, 
	where $\bm{F}_{\mathrm{O}}\in \mathbb{R} ^{3M \times 3M}$ denotes the FIM from observation (i.e., the received signals $\bm{Y}$ shown in (\ref{Y})), and $\bm{F}_{\mathrm{P}}\in \mathbb{R} ^{3M \times 3M}$ denotes the FIM from prior information (i.e., $p_{\mathrm{Z}}(\boldsymbol{\zeta})$ shown in (\ref{p_z})). In the following, we derive $\bm{F}_{\mathrm{O}}$ and $\bm{F}_{\mathrm{P}}$, respectively, based on which the PCRB can be further derived. 
	\subsection{Derivation of FIM from Observation, $\bm{F}_{\mathrm{O}}$}
	$\bm{F}_{\mathrm{O}}$ can be expressed as
\vspace{-0.1cm}
	\begin{align}
		\bm{F}_{\mathrm{O}}=&\mathbb{E}_{\boldsymbol{Y},\boldsymbol{\zeta}}\Big[ \frac{\partial \ln ( f( \boldsymbol{Y}| \boldsymbol{\zeta}))}{\partial \boldsymbol{\zeta}}\Big(\frac{\partial \ln ( f( \boldsymbol{Y}| \boldsymbol{\zeta}))}{\partial \boldsymbol{\zeta}}\Big)^H\Big]\nonumber\\
=&[\bm{F}^{\mathrm{\bm{\theta\theta}}}_{\mathrm{O}},\bm{F}^{\mathrm{\bm{\theta\alpha}}}_{\mathrm{O}};	{\bm{F}^{\mathrm{\bm{\theta\alpha}}}_{\mathrm{O}}}^H,\bm{F}^{\mathrm{\bm{\alpha\alpha}}}_{\mathrm{O}}],
\end{align}
	where $f(\boldsymbol{Y}|\boldsymbol{\zeta})$ denotes the conditional PDF of $\bm{Y}$ given $\bm{\zeta}$, $\bm{F}^{\mathrm{\bm{\theta\theta}}}_{\mathrm{O}}\in \mathbb{R}^{M\times M}$, $\bm{F}^{\mathrm{\bm{\theta\alpha}}}_{\mathrm{O}}\in \mathbb{R}^{M\times2M}$, and $\bm{F}^{\mathrm{\bm{\alpha\alpha}}}_{\mathrm{O}}\in \mathbb{R}^{2M\times 2M}$. 
	
	For $\bm{F}^{\mathrm{\bm{\theta\theta}}}_{\mathrm{O}}$, the $\!(\!m,\!n\!)$-th element with $m\!\neq \! n\!$ can be derived as
	\vspace{-0.2cm}
	\begin{equation}
		[\!\bm{F}^{\mathrm{\bm{\theta\theta}}}_{\mathrm{O}}]_{\!m,\!n\!}\!\!=\!\!\frac{2L}{\sigma^2}\mathfrak{Re}\{\mathbb{E}_{\bm{\zeta}}[\alpha _{m}^{*}\alpha_n\mathrm{tr}(\dot{\boldsymbol{M}}^H\!(\theta_m)\dot{\boldsymbol{M}}(\theta_n)\boldsymbol{R}_X)]\}\!\!=\!\!0,\!\!\!
	\vspace{-0.2cm}
	\end{equation}
	where $\boldsymbol{M}\!(\!\theta_m)\!\!\overset{\Delta}{=}\!\!\boldsymbol{b}( \theta _m) \boldsymbol{a}^H(\theta _m) $ and $\dot{\boldsymbol{M}}(\theta_m)$ denotes the derivative of $\boldsymbol{M}(\theta _m)$. The $(m,m)$-th element can be derived as
	\begin{align}
		&[\bm{F}^{\mathrm{\bm{\theta\theta}}}_{\mathrm{O}}]_{m,m}\!\!=\!\!\frac{2L}{\sigma^2}\mathfrak{Re}\{\mathbb{E}_{\bm{\zeta}}[({\alpha_{m}^{\mathrm{R}}}^2\!\!+\!{\alpha_{m}^{\mathrm{I}}}^2)\mathrm{tr}(\dot{\boldsymbol{M}}^H\!\!(\theta_m)\dot{\boldsymbol{M}}(\theta_m)\boldsymbol{R}_X)]\}\nonumber \\
		&=\frac{2L}{\sigma^2}\mathfrak{Re}\Big\{\Big(\int{{\alpha _{m}^{\mathrm{R}}} ^2p_{\alpha _m}( \alpha _{m}^{\mathrm{R}}) d\alpha _{m}^{\mathrm{R}}}\!+\!\int{{\alpha _{m}^{\mathrm{I}}} ^2p_{\alpha _m}( \alpha _{m}^{\mathrm{I}}) d\alpha _{m}^{\mathrm{I}}}\Big)\nonumber\\
		&\times \mathrm{tr}(\mathbb{E}_{\bm{\theta}}[\dot{\boldsymbol{M}}^H\!( \theta _m) \dot{\boldsymbol{M}}( \theta _m)] \boldsymbol{R}_X)\Big\}\!=\!\frac{4L\sigma _{\alpha _m}^{2}}{\sigma ^2}\mathrm{tr}( \boldsymbol{A}_m\boldsymbol{R}_X),\!\!\!
	\end{align}
\vspace{-0.4cm}%
\par\noindent%
	where $\boldsymbol{A}_m	=\int{\dot{\boldsymbol{M}}^H(\theta _m) \dot{\boldsymbol{M}}(\theta _m) p_{\bm{\Theta}}(\boldsymbol{\theta }) d\boldsymbol{\theta }}\succeq \bm{0}$.
	
	For $\bm{F}^{\mathrm{\bm{\theta\alpha}}}_{\mathrm{O}}$, each $(m,n)$-th $1\times 2$ block is given by
	\begin{align}
		&[\bm{F}^{\mathrm{\bm{\theta\alpha}}}_{\mathrm{O}}]_{m,2n-1:2n}=\frac{2L}{\sigma ^2}\mathfrak{Re}\{ \mathbb{E}_{\bm{\zeta}}[ \alpha _{m}^{*}\mathrm{tr}( \dot{\boldsymbol{M}}^H( \theta _m) \boldsymbol{M}( \theta _n) \boldsymbol{R}_X)\nonumber\\
		\times&[ 1,j]]\}=\frac{2L}{\sigma ^2}\mathfrak{Re} \Big\{ \int{\alpha _{m}^{*}}p_{\alpha _m}\left( \alpha _{m}^{\mathrm{R}} \right) p_{\alpha _m}\left( \alpha _{m}^{\mathrm{I}} \right) d\alpha _{m}^{\mathrm{R}}d\alpha _{m}^{\mathrm{I}} \nonumber\\
		\times&\mathbb{E}_{\bm{\theta}}[\mathrm{tr}(\dot{\boldsymbol{M}}^H(\theta _m) \boldsymbol{M}(\theta _n) \boldsymbol{R}_X)[1,j]] \Big\} =\mathbf{0}_{1\times 2},\quad\forall m,n.
	\end{align}
\vspace{-0.5cm}
	
	For $\bm{F}^{\mathrm{\bm{\alpha\alpha}}}_{\mathrm{O}}$, each $(m,n)$-th $2\times 2$ block is given by
	\vspace{-0.15cm}
	\begin{align}
		&[\bm{F}^{\mathrm{\bm{\alpha\alpha}}}_{\mathrm{O}}]_{2m-1:2m,2n-1,2n}=\frac{2L}{\sigma ^2}\mathfrak{Re}\Big\{\Big[ \begin{matrix}
			1&		j\\
			-j&		1\\
		\end{matrix} \Big]\nonumber\\
		\times& \mathrm{tr}(\mathbb{E}_{\bm{\theta}}[\boldsymbol{M}^H\!(\theta _m) \boldsymbol{M}( \theta _n )] \boldsymbol{R}_X ) \!\Big\} \!\!\ ,\quad \forall m,n.
	\end{align}
\vspace{-0.7cm}%
\par\noindent%

	\subsection{Derivation of FIM from Prior Information, $\bm{F}_{\mathrm{P}}$}
	$\bm{F}_{\mathrm{P}}$ can be expressed as
	\vspace{-0.1cm}
	\begin{equation}
		\label{equ11}
		\bm{F}_{\mathrm{P}}=\mathbb{E} _{\boldsymbol{\zeta }}\Big[ \frac{\partial \ln \left( p_{\mathrm{Z}}\left( \boldsymbol{\zeta } \right) \right)}{\partial \boldsymbol{\zeta }}\big( \frac{\partial \ln \left( p_{\mathrm{Z}}\left( \boldsymbol{\zeta } \right) \right)}{\partial \boldsymbol{\zeta }} \big) ^H \Big],
	\end{equation}
	which can be calculated offline based on $p_{\mathrm{Z}}(\boldsymbol{\zeta})$ in (\ref{p_z}). Particularly, for any $m\neq n$, we have
	\begin{equation}
	\begin{split}
		\left[ \bm{F}_{\mathrm{P}} \right] _{m,n}=&\iint{\frac{\partial \ln \left( p_{\mathrm{Z}}\left( \zeta _m \right) \right)}{\partial \zeta _m}} \frac{\partial \ln \left( p_{\mathrm{Z}}\left( \zeta _n \right) \right)}{\partial \zeta _n} p_{\mathrm{Z}}(\zeta _m)\\
		&\times  p_{\mathrm{Z}}\left( \zeta _n \right) d\zeta _md\zeta _n=0.
	\end{split}
\vspace{-0.5cm}
	\end{equation}
\vspace{-0.2cm}
\par\noindent
	Hence, $\bm{F}_{\mathrm{P}}$ is a diagonal matrix.
	
	\subsection{Derivation of PCRB of MSE for Sensing $\bm{\theta}$}
\addtolength{\topmargin}{-.1in}
	Based on the above, the MSE matrix for estimating $\boldsymbol{\zeta}$ denoted by $\bm{E}$ can be lower bounded by the inverse of the overall FIM $\bm{F}$, i.e., $\bm{E}\succeq \bm{F}^{-1}$. $\bm{F}^{-1}$ is given by
	\begin{equation}
		\bm{F}^{-1}=\Big[ \begin{matrix}
			\big( \bm{F}^{\mathrm{\bm{\theta\theta}}}_{\mathrm{O}}+\left[ \bm{F}_{\mathrm{P}} \right] _{1:M,1:M} \big) ^{-1}&		\boldsymbol{C}\\
			\boldsymbol{C}^H&		\boldsymbol{D}\\
		\end{matrix} \Big],
	\end{equation}
	where $\boldsymbol{C}\in \mathbb{C} ^{M\times 2M}$ and $\boldsymbol{D}\in \mathbb{C} ^{2M\times 2M}$ are functions of $\bm{F}$. Thus, the PCRB for the MSE in estimating $\theta_m$ is given by
	\begin{equation}
	\begin{split}
		\mathrm{PCRB}_{\theta _m}=&\big( \frac{4L\sigma _{\alpha _m}^{2}}{\sigma ^2}\mathrm{tr}( \boldsymbol{A}_m\boldsymbol{R}_X) +[\bm{F}_{\mathrm{P}}] _{m,m} \big) ^{-1} \\
		=&\left( \beta _m\mathrm{tr}( \boldsymbol{A}_m\boldsymbol{R}_X \right) \!+\!\delta _m ) ^{-1},\quad m=1,...,M, \label{PCRB_theta_m}
	\end{split} 
	\end{equation}
	where $\beta _m\!\!\overset{\Delta}{=}\!\!\frac{4L\sigma _{\alpha _m}^{2}}{\sigma ^2}\!\!>\!\!0$ and $\delta _m\!\!\overset{\Delta}{=}\!\!\left[ \bm{F}_{\mathrm{P}} \right] _{m,m}$. Note that in $\mathrm{PCRB}_{\theta _m}$, $\beta _m$, $\bm{A}_m$, and $\delta_m$ are all constant parameters determined by the known prior distribution of $\bm{\zeta}$, thus can be efficiently obtained offline. On the other hand, the only design variable in $\mathrm{PCRB}_{\theta _m}$ is the transmit sample covariance matrix $\bm{R}_X$. Note that based on $\mathrm{PCRB}_{\theta_m}$'s derived in (\ref{PCRB_theta_m}), various multi-target sensing performance metrics can be quantified. In this paper, we adopt the PCRB for the \emph{sum MSE} of estimating the angles of all $M$ targets as the performance metric, which is given by $\sum_{m=1}^M\mathrm{PCRB}_{\theta _m}$.
	
	\section{Problem Formulation}
	Our objective is to minimize the PCRB for the sum MSE in estimating the angles of $M$ targets via optimizing the transmit sample covariance matrix $\bm{R}_X$ under a transmit power budget $P$. The optimization problem is formulated as
	\vspace{-0.1cm}
	\begin{align}
		\mbox{(P1)}\quad \min_{\boldsymbol{R}_X}\quad&\sum_{m=1}^M(\beta _m\mathrm{tr}\left( \boldsymbol{A}_m\boldsymbol{R}_X \right) +\delta _m ) ^{-1}\label{P1obj}\\
		\textrm{s.t.}\quad &\mathrm{tr}(\boldsymbol{R}_X) \leq P\\
					&\boldsymbol{R}_X\succeq \mathbf{0}.
	\end{align}
	\vspace{-0.5cm}
	
	Note that Problem (P1) can be shown to be a convex optimization problem, since $\beta_m>0$ and $\bm{A}_m\succeq \bm{0}$. In the following, we study the structure of the optimal solution to unveil the fundamental transmit signal design principle for multi-target sensing exploiting prior information. Moreover, we will devise a low-complexity algorithm for obtaining the optimal solution to Problem (P1). 
	
	\section{Optimal Solution Structure to Problem (P1)}
	To transform Problem (P1) into a more tractable form, we introduce an auxiliary variable $\boldsymbol{t}=[t_1,... ,t_M]^T\in \mathbb{R} ^M$. By leveraging the Schur complement, Problem (P1) can be shown to be equivalent to the following problem:
	\begin{align}
		\label{equ43}
		\mbox{(P2)}\quad\min_{\boldsymbol{R}_X, \boldsymbol{t }}\quad&\sum_{m=1}^M{t_m}\\
		\textrm{s.t.}\quad&\Big[ \begin{matrix}
			\beta _m\mathrm{tr}( \boldsymbol{A}_m\boldsymbol{R}_X ) +\delta _m&		1\\
			1&		t_m\\
		\end{matrix} \Big] \succeq \mathbf{0},\   m=1,...,\!M\label{equ44}\\
		&\mathrm{tr}\left( \boldsymbol{R}_X \right) \leq P\label{equ45}\\
		&\boldsymbol{R}_X\succeq \mathbf{0}.\label{equ46}
	\end{align}
	Specifically, for any feasible solution of $\mbox{(P2)}$ denoted as $(\boldsymbol{R}_{X}^{\prime},\boldsymbol{t}^{\prime})$, $\boldsymbol{R}_{X}^{\prime}$ is a feasible solution to $\mbox{(P1)}$ which achieves the same objective value of (P1) as that of (P2) with $(\boldsymbol{R}_{X}^{\prime},\boldsymbol{t}^{\prime})$; for any feasible solution of (P1) denoted as $\boldsymbol{R}_{X}^{\prime}$, $(\boldsymbol{R}_{X}^{\prime},\boldsymbol{t}^{\prime})$ with $t_m'=(\beta _m\mathrm{tr}(\boldsymbol{A}_m\boldsymbol{R}_X') +\delta _m ) ^{-1},\forall m$ is a feasible solution to (P2) and achieves the same objective value of (P2) as that of (P1) with $\boldsymbol{R}_{X}^{\prime}$. Particularly, if $(\boldsymbol{R}_{X}^{\star},\boldsymbol{t}^{\star})$ is an optimal solution to $\mbox{(P2)}$, $\boldsymbol{R}_{X}^{\star}$ is also an optimal solution to $\mbox{(P1)}$.
	Problem $\mbox{(P2)}$ is convex and satisfies the Slater's condition \cite{boyd2004convex}. Thus, strong duality holds for (P2). 
	
	In the following, we unveil the structure of the optimal $\boldsymbol{R}_{X}^{\star}$ by investigating (P2) via the Lagrange duality theory. Denote $\boldsymbol{Z}=[\boldsymbol{Z}_1,...,\boldsymbol{Z}_M]$ with $\bm{Z}_m\succeq \bm{0},\forall m$, $\mu\ge 0$, and $\boldsymbol{\varPsi}\succeq \bm{0}$ as the dual variables associated with the constraints in (\ref{equ44}), (\ref{equ45}), and (\ref{equ46}), respectively. The Lagrangian of $\mbox{(P2)}$ is given by
	\begin{align}
		&\mathcal{L}( \boldsymbol{R}_X,\boldsymbol{t},\boldsymbol{Z},\mu ,\boldsymbol{\varPsi })\\ =&\sum_{m=1}^M{t_m}-\sum_{m=1}^M{\mathrm{tr}\left( \boldsymbol{Z}_m\boldsymbol{P}_m \right)}
		+\mu ( \mathrm{tr}( \boldsymbol{R}_X ) -P ) -\mathrm{tr}( \boldsymbol{\varPsi R}_X),\nonumber 
	\end{align}
	where $\boldsymbol{P}_m\overset{\Delta}{=}[\beta_m\mathrm{tr}(\boldsymbol{A}_m\boldsymbol{R}_X)+\delta _m,1;1,t_m]$. By further defining $\boldsymbol{Z}_m=[z_{1,m}, z_{2,m}; z_{2,m}^{*}, z_{3,m}]$, $\mathrm{tr}( \boldsymbol{Z}_m\boldsymbol{P}_m)$ can be written as $\mathrm{tr}\left( \boldsymbol{Z}_m\boldsymbol{P}_m \right) =z_{1,m}\beta _m\mathrm{tr}\left( \boldsymbol{A}_m\boldsymbol{R}_X \right) +z_{1,m}\delta _m+z_{2,m}+z_{2,m}^{*}+z_{3,m}t_m$. Let $\boldsymbol{R}^\star_X,\boldsymbol{t}^\star,\boldsymbol{Z}^\star,\mu^\star,\boldsymbol{\varPsi}^\star$ denote the optimal primal and dual variables. The Karush-Kuhn-Tucker (KKT) optimality conditions consist of (\ref{equ44})-(\ref{equ46}) and
	\begin{align}
		1-z^\star_{3,m}&=0\label{equ47}\\[-3mm]
		-\sum_{m=1}^M{z^\star_{1,m}\beta _m\boldsymbol{A}_{m}^{T}}-\boldsymbol{\varPsi }^\star+\mu^\star \boldsymbol{I}_{N_t}&=\mathbf{0}\label{equ48}\\[-3mm]
		\mathrm{tr}(\boldsymbol{Z}^\star_m\boldsymbol{P}_m^\star) &=0\label{equ49}\\
		\mu^\star ( \mathrm{tr}( \boldsymbol{R}_X^\star) -P) &=0\label{equ50}\\
		\mathrm{tr}(\boldsymbol{\varPsi}^\star\bm{R}_X^\star)&=0\label{equ51}\\
		\mu^\star \geq 0,\, \boldsymbol{\varPsi }^\star\succeq \mathbf{0},\, \boldsymbol{Z}^\star_m\succeq \mathbf{0},&\ \forall m.
	\end{align}

	It follows from (\ref{equ47}) that $z^\star_{3,m}=1$, which indicates $\boldsymbol{Z}^\star_m\ne \mathbf{0}$. Since $\boldsymbol{P}^\star_m\ne \mathbf{0}$, both $\boldsymbol{Z}^\star_m$ and $\boldsymbol{P}^\star_m$ should be singular such that (\ref{equ49}) holds. Therefore, we have $z^\star_{1,m}-| z^\star_{2,m}| ^2{z^\star_{3,m}}^{-1}=0$, and thus $z^\star_{1,m}=| z^\star_{2,m}| ^2\ge 0$. With $\boldsymbol{A}_m\succeq \mathbf{0}$ and $\boldsymbol{\varPsi }^\star\succeq \mathbf{0}$, we have $\mu^\star \ne 0$, thus $\mathrm{tr}\left( \boldsymbol{R}^\star_X \right) =P$. Based on (\ref{equ49}), we further have
	\begin{equation}
		\!\!\!\!t^\star_m=-( \mathrm{tr}( z^\star_{1,m}\beta _m\boldsymbol{A}_m\boldsymbol{R}^\star_X ) +z^\star_{1,m}\delta _m+z^\star_{2,m}+{z^{\star^*}_{2,m}}).\!\!
	\end{equation}
	Define $\boldsymbol{H}^\star_m=z^\star_{1,m}\beta _m\boldsymbol{A}_m$. The optimal solution to $\mbox{(P2)}$ can be obtained by solving the following problem:
	\begin{align}
		\label{equ53}
		\mbox{(P2-I)}\quad\max_{\boldsymbol{R}_X\succeq \mathbf{0}}\quad&\mathrm{tr}\Big(\big( \sum_{m=1}^M\boldsymbol{H}^\star_m\big)\boldsymbol{R}_X \Big)\\
		\textrm{s.t.}\quad&\mathrm{tr}\left( \boldsymbol{R}_X \right) \leq P.
	\end{align}
	Based on (P2-I), we have the following proposition.
	\begin{proposition}
		For Problem (P1), an optimal transmit sample covariance matrix is $\boldsymbol{R}_{X}^{\star}=\sum_{n=1}^N{P_n\boldsymbol{q}_n\boldsymbol{q}_{n}^{H}}$, where $\sum_{n=1}^N{P_n}=P$ and $\boldsymbol{q}_n$ is the eigenvector corresponding to the strongest eigenvalue of $\sum_{m=1}^M{\boldsymbol{H}^\star_m}$. 
	\end{proposition}
	\begin{IEEEproof}
		Denote the eigenvalue decomposition (EVD) of $\sum_{m=1}^M{\boldsymbol{H}_m}^\star$ as $\boldsymbol{Q\varLambda Q}^H$, where $\bm{Q}=[\bm{q}_1,...,\bm{q}_{N_t}]$ and $\boldsymbol{\varLambda}=\mathrm{diag}\{\lambda _1,... ,\lambda _{N_t} \}$ with $\lambda _1=...=\lambda _N>\lambda _{N+1}\ge ...\ge \lambda _{N_t}$. The optimal solution to (P2-I) can be shown to have the structure of $\boldsymbol{R}_{X}^{\star}=\sum_{n=1}^N{P_n\boldsymbol{q}_n\boldsymbol{q}_{n}^{H}}$ with $\sum_{n=1}^N{P_n}=P$. Proposition 1 thus follows based on the equivalence between (P1) and (P2).
	\end{IEEEproof}

	Proposition 1 indicates that the optimal transmit sample covariance matrix design is dependent on $\{\bm{A}_m\}_{m=1}^M$ which is determined by the joint PDF of all angles, $p_{\bm{\Theta}}(\bm{\theta})$, the array steering vectors at the BS transmit and receive antennas (i.e., the antenna layouts), as well as other system parameters including the transmit power budget, $P$, total number of samples used for sensing, $L$, and variances of the overall reflection coefficients for each target, $\{\sigma_{\alpha_m}^2\}_{m=1}^M$.
	
	In the following, we propose effective algorithms for obtaining $\bm{R}_X^\star$ and analyze their computational complexities.
	
	\section{Optimal Solution to Problem (P1) via Interior-Point Method for Semi-Definite Program (SDP)} \label{4C}
	Problem $\mbox{(P2)}$ is a standard SDP that can be solved via the interior-point method \cite{boyd2004convex}, for which the details are omitted for brevity. The complexity for the interior-point method is analyzed as follows. First, the complexity for obtaining $\{\boldsymbol{A}_m\}_{m=1}^M$ and $\{\delta_m\}_{m=1}^M$ is given by $\mathcal{O}(MN_{t}^{2}\varpi)$, where $\varpi $ denotes the number of flops for performing one-dimensional integration. Next, since $\mbox{(P2)}$ is an SDP consisting of $2N_{t}^{2}+M$ real variables, the interior-point method requires at most $\mathcal{O} ( \sqrt{2N_{t}^{2}+M} )$ iterations to converge, and the complexity per iteration is $\mathcal{O} ( ( 2N_{t}^{2}+M ) ^3)$ \cite{sidiropoulos2006transmit,boyd2004convex}. Hence, the overall complexity for solving the SDP (P2) and equivalently (P1) via the interior-point method is $\mathcal{O}( MN_{t}^{2}\varpi +( 2N_{t}^2+M )^{3.5} )$.
	
	\section{Optimal Solution to Problem (P1) via Sum-of-Ratios Iterative Algorithm}
	Note that the complexity of the interior-point method for solving the SDP increases rapidly with the number of transmit antennas, $N_t$, and the number of targets, $M$. In this section, we propose an alternative algorithm for obtaining the optimal solution to (P1) with lower complexity. Note that (P1) is fundamentally a \emph{sum-of-ratios} minimization problem, which can be equivalently transformed to the following parametric program \cite{jong2012efficient}:
	\begin{align}
		\mbox{(P3)}\quad\min_{\boldsymbol{R}_X, \boldsymbol{\varphi }} \quad&\sum_{m=1}^M{\varphi _m}\\
		\textrm{s.t.}\quad&\left( \beta _m\mathrm{tr}\left( \boldsymbol{A}_m\boldsymbol{R}_X \right) +\delta _m \right) ^{-1}\leq \varphi _m,  m=1,...,M\label{equ23}\\
		&\mathrm{tr}\left( \boldsymbol{R}_X \right) \le P\label{equ24}\\
		&\boldsymbol{R}_X\succeq \mathbf{0}, \label{equ25}
	\end{align}
	where $\boldsymbol{\varphi}=[\varphi_1,...,\varphi_M]^T\in \mathbb{R} ^M$ is an auxiliary variable. Note that (P3) is a convex optimization problem which satisfies the Slater's condition, thus strong duality holds. Denote $\boldsymbol{u}=[u_1,...,u_M]^T\succeq \bm{0}$, $\mu\ge 0$, and $\boldsymbol{\varPsi }\succeq \bm{0}$ as the dual variables associated with the constraints in (\ref{equ23}), (\ref{equ24}), and (\ref{equ25}), respectively. The Lagrangian of $\mbox{(P3)}$ is given by
	\begin{align}
		&\mathcal{L}( \boldsymbol{R}_X,\boldsymbol{\varphi },\boldsymbol{u},\mu ,\boldsymbol{\varPsi } ) =\sum_{m=1}^M\varphi _m+u_m( 1-\varphi _m( \beta _m\mathrm{tr}( \boldsymbol{A}_m\boldsymbol{R}_X )\nonumber\\
		&\qquad\qquad +\delta _m ) )+\mu ( \mathrm{tr}( \boldsymbol{R}_X ) -P) -\mathrm{tr}( \boldsymbol{\varPsi R}_X).
	\end{align}
	Let $ \boldsymbol{R}^\star_X,\boldsymbol{\varphi }^\star,\boldsymbol{u}^\star,\mu^\star,\boldsymbol{\varPsi}^\star $ denote the optimal primal and dual variables for (P3). The KKT optimality conditions for $\mbox{(P3)}$ include (\ref{equ23})-(\ref{equ25}) and
	\begin{align}
		-\sum_{m=1}^M{u_m^\star\varphi^\star_m\beta _m\boldsymbol{A}_{m}^{T}}+\mu^\star \boldsymbol{I}_{N_t}-\boldsymbol{\varPsi }^\star&=\mathbf{0}\label{equ26}\\
		1-u_m^\star\left( \beta _m\mathrm{tr}\left( \boldsymbol{A}_m\boldsymbol{R}^\star_X \right) +\delta _m \right) &=0\label{equ27}\\
		u_m^\star[ 1-\varphi^\star_m( \beta _m\mathrm{tr}( \boldsymbol{A}_m\boldsymbol{R}^\star_X) +\delta _m) ] &=0\label{equ28}\\
		\mu^\star ( \mathrm{tr}( \boldsymbol{R}^\star_X ) -P )& =0 \\
           \mathrm{tr}(\boldsymbol{\varPsi}^\star\bm{R}^\star_X)&=0 \label{equ29}\\
		\mu^\star \geq 0,\ \boldsymbol{\varPsi }^\star\succeq \mathbf{0},\ u_m^\star\geq 0,&\ \forall m.\label{equ31}
	\end{align}
	Since $\beta _m\mathrm{tr}(\boldsymbol{A}_m\boldsymbol{R}^\star_X) +\delta _m>0$, (\ref{equ27}) is equivalent to
	\begin{equation}
		\label{equ37}
		u_{m}^{\star}=\left( \beta _m\mathrm{tr(}\boldsymbol{A}_m\boldsymbol{R}_{X}^{\star})+\delta _m \right) ^{-1},\quad  m=1,... ,M.
	\end{equation}
	Hence, (\ref{equ28}) is equivalent to
	\begin{equation}
		\label{equ38}
		1-\varphi^\star_m( \beta _m\mathrm{tr}( \boldsymbol{A}_m\boldsymbol{R}^\star_X ) +\delta _m ) =0,\quad m=1,... ,M.
	\end{equation}
	Therefore, (P3) and equivalently (P1) are equivalent to the following problem:
	\vspace{-0.3cm}
	\begin{align}
		\label{equ39}
		\mbox{(P3-eqv)}\quad\min_{\boldsymbol{R}_X}\quad&\sum_{m=1}^M\!{u_m^\star[ 1\!\!-\!\!\varphi^\star_m( \beta _m\mathrm{tr}( \boldsymbol{A}_m\boldsymbol{R}_X )\! +\!\delta _m ) ]}\\
		\textrm{s.t.}\quad  &\mathrm{tr}( \boldsymbol{R}_X ) \leq P\\
		&\boldsymbol{R}_X\succeq \mathbf{0}.
	\end{align}
\vspace{-0.55cm}%
\par\noindent%
	Note that the sum-of-ratios problem $\mbox{(P1)}$ is equivalently transformed to a parametric optimization problem (P3-eqv) with an objective function in subtractive form.

In the following, we obtain the optimal solution to (P3-eqv) via iteratively optimizing $\bm{R}_X$ and $(\boldsymbol{\varphi},\boldsymbol{u})$. Firstly, with given $(\boldsymbol{\varphi},\boldsymbol{u})$, (P3-eqv) is equivalent to the following problem:
	\vspace{-0.2cm}
	\begin{align}
		\label{equ42}
		\mbox{(P3-eqv')}\quad \max_{\boldsymbol{R}_X\succeq \mathbf{0}}\quad &\mathrm{tr}( \sum_{m=1}^M{u_m\varphi _m\beta _m\boldsymbol{A}_m}\boldsymbol{R}_X )\\
		\textrm{s.t.}\quad &\mathrm{tr}( \boldsymbol{R}_X ) \leq P.
	\end{align}
\vspace{-0.4cm}%
\par\noindent%
	Denote $\sum_{m=1}^M{\!\!u_m\varphi _m\beta _m\boldsymbol{A}_m}\!\!=\!\tilde{\boldsymbol{Q}}\tilde{\boldsymbol{\varLambda}}\tilde{\boldsymbol{Q}}^H$, where $\tilde{\bm{Q}}\!\!=\!\![\tilde{\bm{q}}_1,...,\tilde{\bm{q}}_{N_t}]$ and $\tilde{\boldsymbol{\varLambda}}=\mathrm{diag}\{ \tilde{\lambda}_1,...,\tilde{\lambda}_{N_t} \}$ with $\tilde{\lambda}_1=...=\tilde{\lambda}_N>\tilde{\lambda}_{N+1}\ge ...\ge \tilde{\lambda}_{N_t}$. It can be shown that $\tilde{\boldsymbol{R}}_{X}^{\star}=\sum_{n=1}^N{\tilde{P}_n\tilde{\boldsymbol{q}}_n\tilde{\boldsymbol{q}}_{n}^{H}}$ with $\sum_{n=1}^N{\tilde{P}_n}=P$ is an optimal solution to (P3-eqv') which can be obtained via EVD. Then, with given $\bm{R}_X$, $( \boldsymbol{\varphi },\boldsymbol{u})$ can be updated via the modified Newton's method \cite{jong2012efficient}. By iteratively updating $\bm{R}_X$ and $( \boldsymbol{\varphi },\boldsymbol{u})$ until convergence, the \emph{optimal solution} to (P3) and equivalently (P1) can be obtained \cite{jong2012efficient}. The detailed steps of the proposed sum-of-ratios iterative algorithm are summarized in Algorithm 1.
	
	In Algorithm 1, the complexity of obtaining the optimal solution to $\mbox{(P3-eqv')}$ is dominated by the EVD operation, which has a complexity of $\mathcal{O} ( N_{t}^{3} )$. The update of $( \boldsymbol{\varphi },\boldsymbol{u})$ in each $k$-th iteration requires $\mathcal{O} ( ( i_k+1 ) M )$ flops, where $i_k$ is defined in Algorithm \ref{alg1}. Suppose the algorithm needs $\tilde{K}$ iterations for convergence, the overall complexity of Algorithm 1 is $\mathcal{O}(MN_{t}^{2}\varpi+\tilde{K}N_{t}^{3}+\sum_{k=1}^{\tilde{K}}{\!(i_k\!+\!1\!)M} )$, which is in a lower order of the number of transmit antennas and the number of targets compared to that of the interior-point method for SDP.
	
	\begin{algorithm}[htbp]
		\renewcommand{\algorithmicrequire}{\textbf{Input:}}
		\renewcommand{\algorithmicensure}{\textbf{Output:}}
		\caption{Proposed Sum-of-Ratios Iterative Algorithm for Obtaining the Optimal Solution to Problem (P1)}
		\label{alg1}
		\begin{algorithmic}[1]
			\REQUIRE $p_{\bm{\Theta}}(\bm{\theta})$, $P$, $\bm{M}(\theta)$, $\{\beta_m\}_{m=1}^M$,  convergence tolerance parameters $\mathbf{\Delta }\in \mathbb{R} ^{2M\times 1}$, $\xi \in ( 0,1 )$, and $\varepsilon \in ( 0,1 )$.
                \STATE Initialize $\varphi _{m}^{0}\!\!=\!\!\frac{1}{P\beta_mtr(A_m)/N_t+\delta_m}$, $\!u_{m}^{0}\!\!=\!\varphi _{m,}^{0}$, $\!\forall m$, and $k\!\!=\!0$.
			\STATE With given $\boldsymbol{\eta }^k=( \boldsymbol{\varphi }^k,\boldsymbol{u}^k )$, solve (P3-eqv') via EVD and obtain the optimal solution $\boldsymbol{R}_{X}^{k}$.
			\STATE With given $\bm{R}_X^k$, set $\varDelta _m( \boldsymbol{\eta }^k ) \!=\!-1+\varphi _{m}^{k}( \beta _m\mathrm{tr}( \boldsymbol{A}_m\boldsymbol{R}_{X}^{k} ) +\delta _m )$, and $\varDelta _{M+m}( \boldsymbol{\eta }^k ) =-1+u_{m}^{k}( \beta _m\mathrm{tr}( \boldsymbol{A}_m\boldsymbol{R}_{X}^{k} ) +\delta _m )$.
			\IF{$\varDelta ( \boldsymbol{\eta }^k ) \preceq \mathbf{\Delta }$}
			\STATE $\boldsymbol{R}_{X}^{\star}=\boldsymbol{R}_{X}^{k}$ is the optimal solution to (P1). Stop.
			\ELSE
			\STATE Let $i_k$ denote the smallest non-negative integer satisfying $\| \varDelta ( \boldsymbol{\eta }^k+\xi ^i\boldsymbol{p}^k ) \| \leq ( 1-\varepsilon \xi ^i )\| \varDelta ( \boldsymbol{\eta }^k ) \|$, where $\boldsymbol{p}^k=-[ \varDelta ^{\prime}( \boldsymbol{\eta }^k ) ] ^{-1}\varDelta ( \boldsymbol{\eta }^k )$ and $\varDelta ^{\prime}( \boldsymbol{\eta }^k )$ is the Jacobian matrix of $\varDelta ( \boldsymbol{\eta }^k )$.
			\STATE Let $\nu _k=\xi ^{i_k}$, set $\boldsymbol{\eta }^{k+1}=\boldsymbol{\eta }^k+\nu _k\boldsymbol{p}^k$.
			\STATE $k\gets k+1$, go to step 2.
			\ENDIF 
			\ENSURE Optimal transmit sample covariance matrix $\boldsymbol{R}_{X}^{\star}$.
		\end{algorithmic}  
	\end{algorithm}
	\vspace{-0.2cm}
	
	\section{Numerical Results}
\addtolength{\topmargin}{0.2in}
	In this section, we present numerical results to evaluate the performance of our proposed transmit signal designs in multi-target sensing. Consider a MIMO radar system with $N_t=10$ transmit antennas and $N_r=12$ receive antennas under a ULA layout with half-wavelength antenna spacing. The array steering vectors are thus given by $\boldsymbol{a}( \theta _m ) \!=\![ e^{\frac{-\!j\pi ( N_t\!-\!1 )\!\sin\! \theta _m}{2}},\!e^{\frac{-\!j\pi ( N_t\!-\!3 )\!\sin\! \theta _m}{2}},... ,\!e^{\frac{j\pi ( N_t\!-\!1 )\!\sin\! \theta _m}{2}} ] ^T$ and $\boldsymbol{b}( \theta _m ) \!\! =[ e^{\frac{-\!j\pi ( N_r\!-\!1 )\sin \theta _m}{2}},\!e^{\frac{-\!j\pi ( N_r-3 )\sin \theta _m}{2}},... ,e^{\frac{j\pi ( N_r-1 )\sin \theta _m}{2}} ] ^T$. Each target's angle to be sensed lies in the region $[-\frac{\pi}{2},\frac{\pi}{2})$. The number of samples of the transmit probing signal is set as $L\!\!=\!\!25$, and the average noise power at the BS receiver is $\sigma ^2\!=\!-90$ dBm. Unless otherwise specified, we set $M=2$ \footnote{It is worth noting that this work is applicable to any value of $M$.} and $P\!=\!30$ dBm. In practice, each target's angle is typically concentrated around one or multiple nominal angles. Motivated by this, we consider a practical PDF for $\bm{\theta}$, where the angles of different targets are independent of each other, and each angle $\theta_m$ has a PDF following the \emph{Gaussian mixture model} given by $p_{\Theta_m}(\theta _m) =\sum_{i=1}^{K_m}{\frac{p_{m,i}}{\sqrt{2\pi}\sigma _{m,i}}e^{-\frac{( \theta _m-\bar{\theta} _{m,i} ) ^2}{2\sigma _{m,i}^{2}}}},\ \forall m$. $K_m$ denotes the number of Gaussian PDF components of the $m$-th target; $p_{m,i}$, $\bar{\theta} _{m,i}\in [ -\frac{\pi}{2},\frac{\pi}{2} )$, and $\sigma _{m,i}^2$ represent the weight, mean, and variance of the $i$-th Gaussian PDF, respectively. We set $K_1\!\!=\!\!4$, $\bar{\theta} _{1,1}\!\!=\!\!-0.77$, $\bar{\theta} _{1,2}\!=\!-0.4$, $\bar{\theta} _{1,3}\!=\!-0.14$, $\bar{\theta} _{1,4}\!=\!0.82$, $p_{1,1}\!=\!0.17$, $p_{1,2}\!=\!0.33$, $p_{1,3}\!=\!0.15$, $p_{1,4}\!=\!0.35$, $\sigma_{1,1}^2\!\!=\!\!\sigma_{1,2}^2\!\!=\!\!\sigma_{1,3}^2\!\!=\!\!10^{-2}$, $\sigma_{1,1}^2\!\!=\!\!10^{-2.5}$; and $K_2\!\!=\!\!4$, $\bar{\theta} _{2,1}\!\!=\!\!-0.73$, $\bar{\theta} _{2,2}\!\!=\!\!-0.45$, $\bar{\theta} _{2,3}\!\!=\!\!0.76$, $\bar{\theta} _{2,4}\!\!=\!\!0.8$, $p_{2,1}\!\!=\!\!0.16$, $p_{2,2}\!\!=\!\!0.39$, $p_{2,3}\!\!=\!\!0.25$, $p_{2,4}\!\!=\!\!0.2$, $\sigma_{2,1}^2\!\!=\!\!\sigma_{2,3}^2\!\!=\!\!10^{-2.5}$, $\sigma_{2,2}^2\!\!=\!\!\sigma_{2,4}^2\!\!=\!\!10^{-2}$. We further set $\sigma _{\alpha _m}^{2}\!\!=\!\!10^{-11}$ for all $m$, $\xi \!\!=\!\!0.5$, $\varepsilon \!\!=\!\!0.5$, and $\Delta_n\!\!=\!\!10^{-3},\forall n$ for the proposed sum-of-ratios iterative algorithm.
	
\begin{figure}[t]
		\centerline{\includegraphics[width=0.35\textwidth]{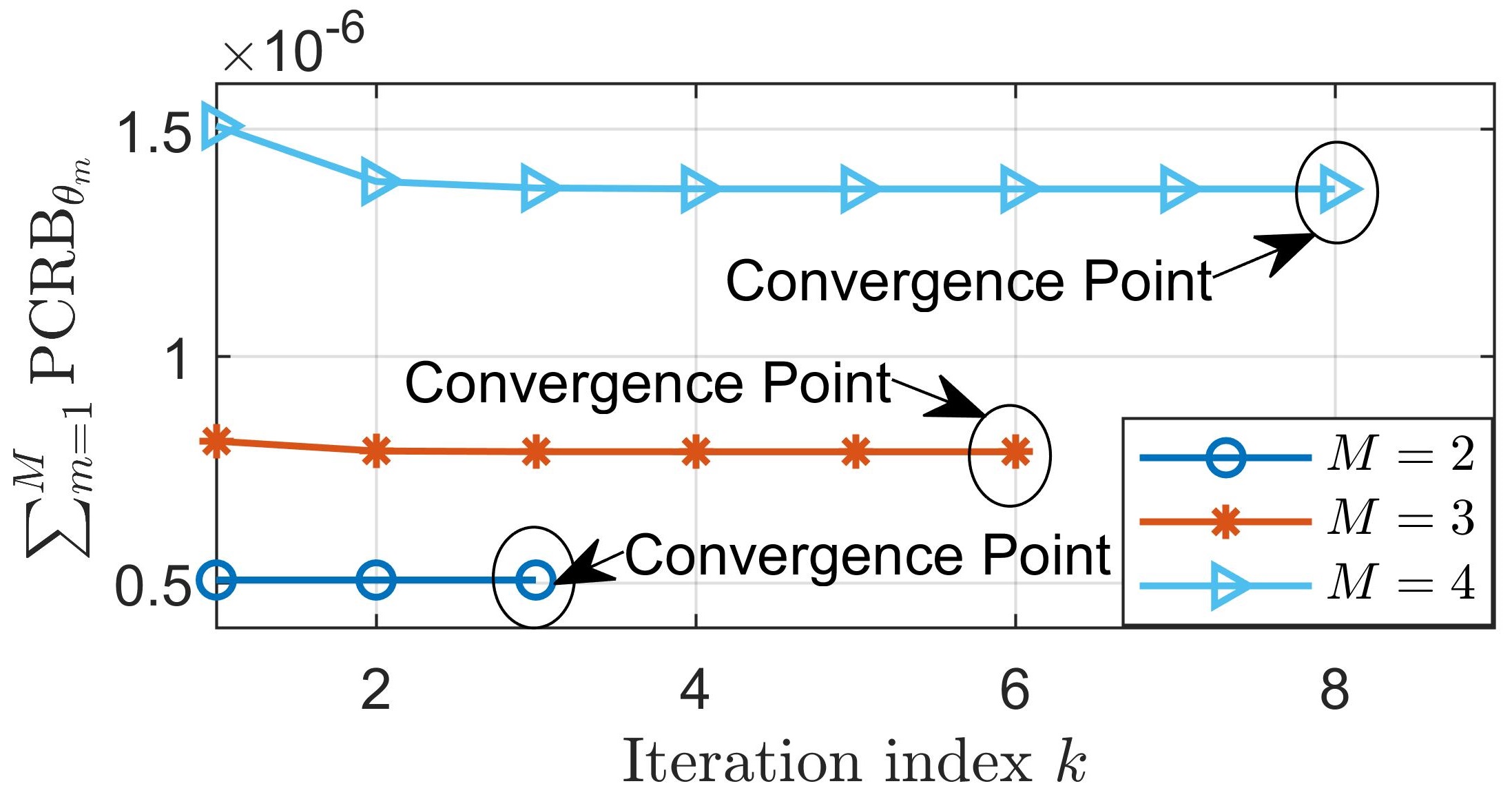}}
		\vspace{-0.4cm}
		\caption{Convergence behavior of Algorithm 1.}
		\vspace{-0.5cm}
		\label{fig1}
	\end{figure}

\begin{figure}[t]
		\vspace{0.05cm}
		\centerline{\includegraphics[width=0.38\textwidth]{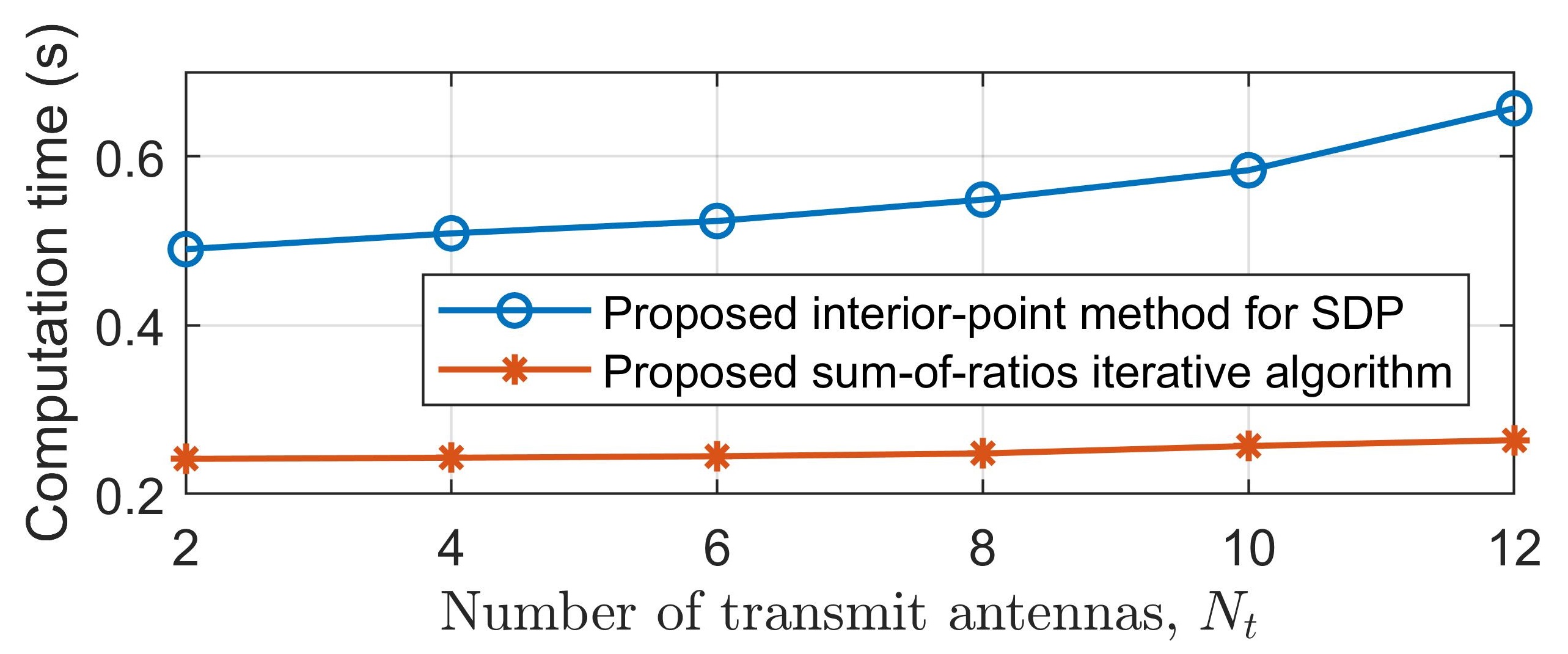}}
		\vspace{-0.41cm}
		\caption{Computation time of solving $\mbox{(P1)}$ using different algorithms.}
		\label{fig4}
		\vspace{-7mm}
	\end{figure}

	First, we show in Fig. \ref{fig1} the sum of $\mathrm{PCRB}_{\theta _m}$ versus the iteration index $k$ of the proposed sum-of-ratios iterative algorithm under different $M$.\footnote{For the case of $M=3$ and $M=4$, we set $K_3\!\!=\!\!4$, $\bar{\theta} _{3,1}\!\!=\!\!-1.2$, $\bar{\theta} _{3,2}\!\!=\!\!-0.5$, $\bar{\theta} _{3,3}\!\!=\!\!0.5$,$\bar{\theta} _{3,4}\!\!=\!\!0.75$, $p_{3,1}\!\!=\!\!0.2$, $p_{3,2}\!\!=\!\!0.35$, $p_{3,3}\!\!=\!\!0.2$,$p_{3,4}\!\!=\!\!0.25$, $\sigma_{3,1}^2\!\!=\!\!\sigma_{3,2}^2\!\!=\!\!\sigma_{3,3}^2\!\!=10^{-3}$, $\sigma_{3,4}^2\!\!=10^{-4}$. For the case of $M=4$, we set $K_4\!\!=\!\!4$, $\bar{\theta} _{4,1}\!\!=\!\!-1.5$, $\bar{\theta} _{4,2}\!\!=\!\!-0.6$, $\bar{\theta} _{4,3}\!\!=\!\!0.9$,$\bar{\theta} _{4,4}\!\!=\!\!1.3$, $p_{4,1}\!\!=\!\!p_{4,3}\!\!=\!\!0.3$, $p_{4,2}\!\!=\!\!p_{4,4}\!\!=\!\!0.2$, $\sigma_{4,1}^2\!\!=\!\!\sigma_{4,2}^2\!\!=\sigma_{4,3}^2\!\!=\!\!\sigma_{4,4}^2\!\!=\!\!10^{-4}$.} It is observed that the proposed sum-of-ratios iterative algorithm converges to the optimal point within ten iterations for all values of $M$, which indicates its fast convergence rate. In Fig. \ref{fig4}, we show the computation time for obtaining the optimal solution to $\mbox{(P1)}$ by both the interior-point method for SDP and the proposed sum-of-ratios iterative algorithm. We use MATLAB on a computer with an Intel Core i7 2.10-GHz CPU and 32 GB of memory. It is observed that the proposed sum-of-ratios iterative algorithm consumes much less computation time than the interior-point method due to fewer iterations and less complex operations in each iteration. Moreover, the computation time gap increases as $N_t$ increases, which is consistent with our results in Sections VI and VII.

		\begin{figure}[t]
		\centerline{\includegraphics[width=0.40\textwidth]{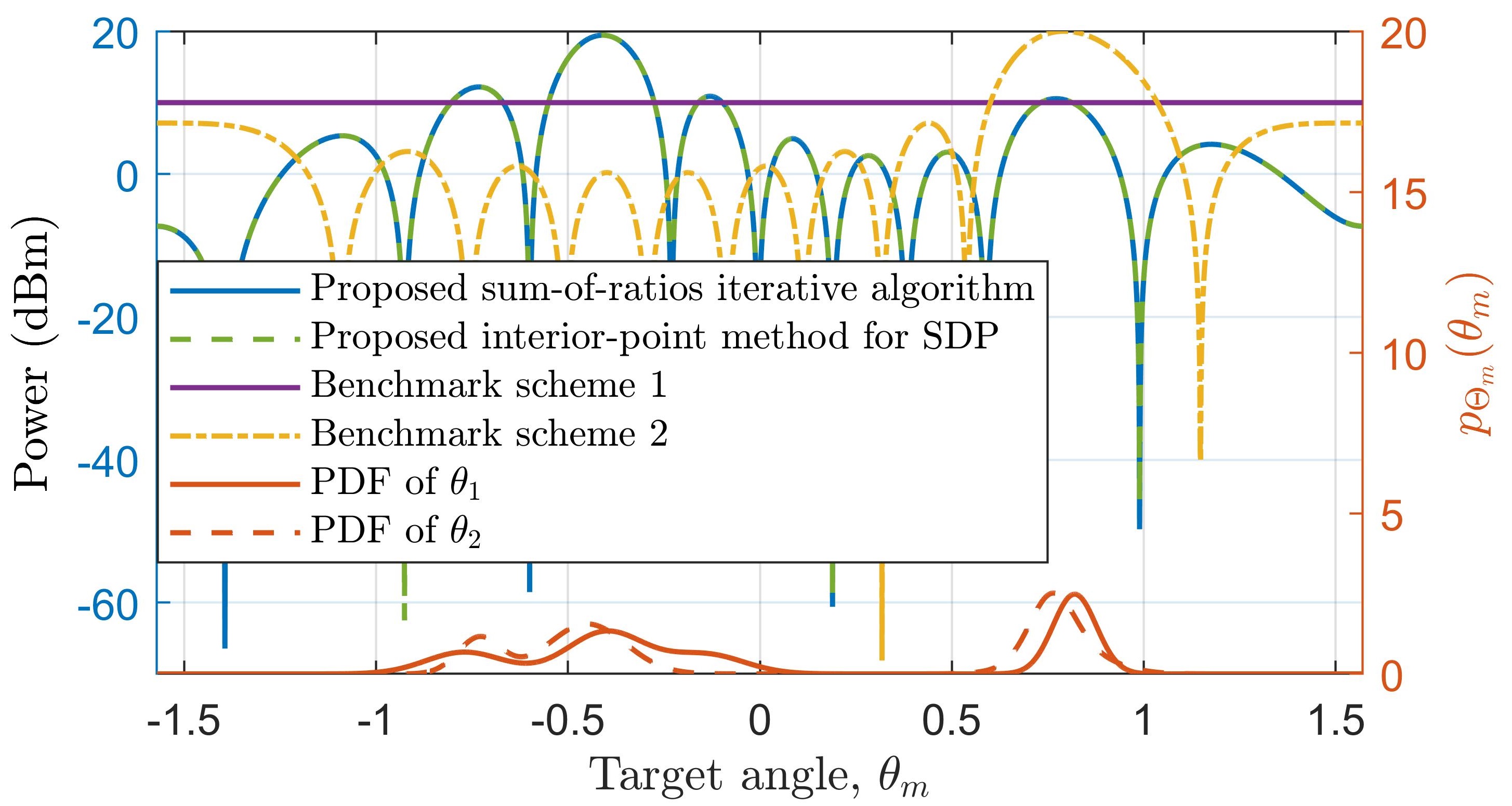}}
		\vspace{-0.3cm}
		\caption{Radiated power pattern and PDF of the targets over different angles.}
		\label{fig_power_pattern}
		\vspace{-0.4cm}
	\end{figure}
	\begin{figure}[t]
	\centerline{\includegraphics[width=0.40\textwidth]{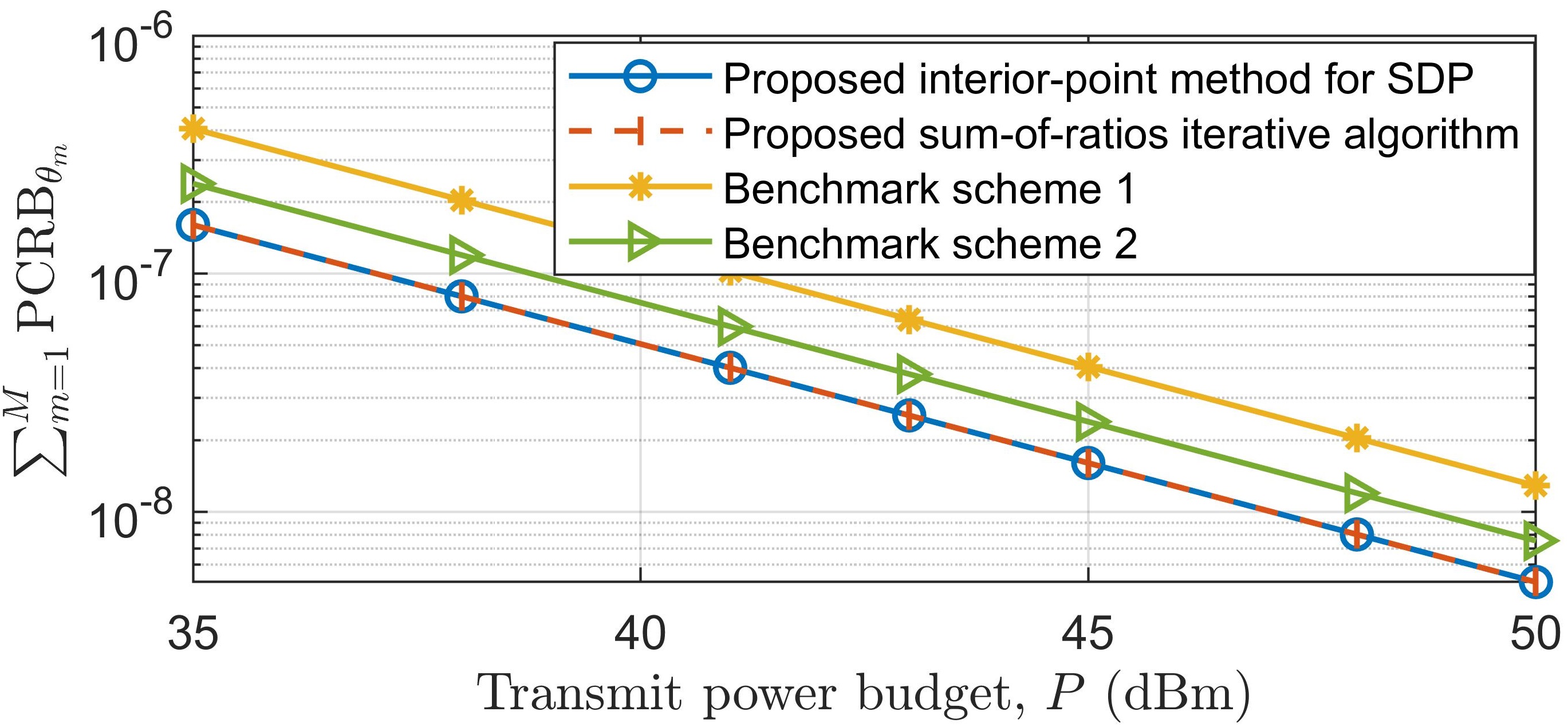}}
	\vspace{-0.3cm}
	\caption{PCRB with different transmit signal designs.}
	\label{fig_PCRB}
	\vspace{-0.6cm}
\end{figure}
	
	Next, we compare the performance of our proposed transmit signal design with two benchmark schemes: 1) {\bf{Benchmark scheme 1}}: Heuristic transmit covariance design with $\boldsymbol{R}_{X}=\frac{P}{N_t}\boldsymbol{I}_{N_t}$; 2) {\bf{Benchmark scheme 2}}: CRB-based transmit covariance design assuming the targets are located at their most probable angles, where $\boldsymbol{R}_{X}$ is designed to minimize the sum of $\mathrm{CRB}_{\theta _m}=( \beta _m\mathrm{tr}( \dot{\boldsymbol{M}}^H( \tilde{\theta} _m ) \dot{\boldsymbol{M}}( \tilde{\theta} _m ) \boldsymbol{R}_X ) ) ^{-1}$, with $\tilde{\theta} _m=\arg\max p_{\Theta_m}(\theta_m),\forall m$.
	
	%
	%

	Fig. \ref{fig_power_pattern} shows the radiated power pattern with different transmit signal designs and the PDF of the targets over different angles. It is observed that our proposed designs are able to achieve sufficient power focusing around every range with high probability densities. Moreover, Fig. \ref{fig_PCRB} shows the PCRB versus the transmit power budget with different transmit signal designs. It is observed that our proposed designs via interior-point method and sum-of-ratios iterative algorithm achieve the same PCRB, and outperform both benchmark schemes.

	\section{Conclusions}
	This paper studied the transmit sample covariance matrix optimization in a MIMO radar system for sensing the unknown and random angles of multiple targets by exploiting their joint distribution. To analytically quantify the multi-target sensing performance, the PCRB for the MSE matrix was derived, which is general for any prior distribution. Then, the transmit covariance matrix optimization problem was formulated to minimize the PCRB for the sum MSE in estimating all targets' angles. Moreover, a novel sum-of-ratios iterative algorithm was proposed which finds the optimal solution with low complexity. Numerical results showed that the proposed algorithm has a fast convergence rate, and outperforms various benchmark schemes in terms of PCRB.

	\bibliographystyle{IEEEtran}
	\bibliography{ref}

\end{document}